\documentclass[preprint,nofootinbib,aps]{revtex4}
 \usepackage{graphicx}
 
\begin{document}
 \title{Charmless $B_{c} \to PP, PV$ decays in the QCD factorization approach}
 \author{Na Wang}
 \email[E-mail address:]{wangna05001@126.com}
 \affiliation{Institute of Particle Physics and Key Laboratory of Quark and Lepton Physics~(MOE), \\
 Central China Normal University, Wuhan, Hubei 430079, P.~R. China}
  \begin{abstract}
  The charmless $B_{c} \to PP, PV$~(where $P$ and $V$ denote the light pseudoscalar and vector mesons, respectively) decays can occur only via the weak annihilation diagrams within the Standard Model and provide, therefore, an ideal place to probe the strength of annihilation contribution in hadronic $B_{u,d,s}$ decays. In this paper, we study these kinds of decays in the framework of QCD factorization, by adopting two different schemes: scheme I is similar to the method usually adopted in the QCD factorization approach, while scheme II is based on the infrared behavior of gluon propagator and running coupling. For comparison, in our calculation, we adopt three kinds of wave functions for $B_{c}$ meson. It is found that: (a) The predicted branching ratios in scheme I are, however, quite small and are almost impossible to be measured at the LHCb experiment. (b) In scheme II, by assigning a dynamical gluon mass to the gluon propagator, we can avoid enhancements of the contribution from soft endpoint region. The strength of annihilation contributions predicted in scheme II is enhanced compared to that obtained in scheme I. However, the predicted branching ratios are still smaller than the corresponding ones obtained in the perturbative QCD approach. The large discrepancies among these theoretical predictions indicate that more detailed studies of these decays are urgently needed and will be tested by the future measurements performed at the LHCb experiment.
 \end{abstract}
 \maketitle

 \section{Introduction}
 \label{sec1}
 The $B_{c}$ meson is the lowest-lying bound state of two heavy quarks with different flavors~($\bar b$ and $c$). Due to its flavor quantum numbers $B=C=\pm1$ and being below the $BD$ threshold, the $B_{c}$ meson is stable against strong and electromagnetic interactions and can decay only via weak interaction. Furthermore, the $B_{c}$ meson has a sufficiently large mass, each of the two heavy quarks can decay individually, resulting in rich decay channels~\cite{Chang:1992pt}. Therefore, the $B_{c}$ meson is an ideal system to study weak decays of heavy mesons~\cite{Brambilla:2004wf}.

 The experimental studies of $B_c$-meson properties started in 1998 when the Collider Detector at Fermilab~(CDF) reported the first observation of $B_c$ meson through the semi-leptonic decay modes $B_c\to J/\Psi\ell^{+}X~(\ell=e,\mu)$~\cite{Abe:1998wi}. Thanks to the fruitful performance of the CDF, D0 and LHCb collaborations, both the mass~\cite{Aaltonen:2007gv,Abazov:2008kv,Aaij:2012dd} and the lifetime~\cite{Abazov:2008rba,Abulencia:2006zu,Aaij:2014bva} of the $B_c$ meson have been measured quite accurately. At the Large Hadron Collider~(LHC) with a luminosity of about ${\cal L}=10^{34}{\rm cm}^{-2}{\rm s}^{-1}$, one could expect around $5\times 10^{10}$ $B_c$ events per year~\cite{Altarelli:2008xy}. In addition, several hadronic $B_{c}$ decay channels, such as $B_{c}^{+}\to J/\psi K^{+}$~\cite{Aaij:2013vcx} and $B_{c}^{+}\to B_{s}^{0}\pi^{+}$~\cite{Aaij:2013cda}, have also been observed for the first time. In the following years, the properties of $B_c$ meson and the dynamics involved in $B_c$ decays will be further exploited through the precision measurements at the LHC with its high collision energy and high luminosity, opening therefore a golden era of $B_c$ physics~\cite{Bediaga:2012py}.

 The theoretical investigations have also been carried out on the properties of $B_c$ meson, such as its lifetime, its decay constant, and some of its form factors, based on different theoretical frameworks~\cite{Brambilla:2004wf}. Due to its heavy-heavy nature and the participation of strong interaction, the hadronic $B_c$ decays are extremely complicated but, at the same time, provide great opportunities to study the perturbative and non-perturbative QCD, and final state interactions in heavy meson decays. Being weakly decaying and doubly heavy flavor meson, it also offers a novel window for studying the heavy-quark dynamics that is inaccessible through the $b\bar{b}$ and $c\bar{c}$ quarkonia~\cite{Brambilla:2004wf}. These features have motivated an extensive study of $B_c$ decays in various theoretical approaches in the literature~\cite{Sun:2008wa}.

 In this paper, we shall focus on the two-body charmless hadronic $B_c$ decays, which can proceed only via the weak annihilation diagrams in the Standard Model~(SM): the initial $\bar{b}$ and $c$ quarks annihilate into $u$ and $\bar{d}/\bar{s}$ quarks, which form two light mesons by hadronizing with a $q\bar{q}$~($q=u,d,s$) pair emitted from a gluon. Detailed studies of these decays will be certainly helpful for further improving our understanding of the weak annihilation contributions, the size of which is currently an important issue in $B$ physics.

 The recent measurements of the $B_{u,d,s}$ decays, especially of the pure annihilation processes $B_s\to \pi^+\pi^-$ and $B_d\to K^+K^-$~\cite{Aaltonen:2011jv,Aaij:2012as}, indicate that the annihilation topologies can be significant, contrary to  the common belief of their power suppression in the heavy-quark limit~\cite{Bauer:1986bm}. Although it was later noticed theoretically that the annihilation amplitudes may not be negligibly small in realistic $B$-meson decays~\cite{Keum:2000ph}, it is still very hard to make a reliable calculation of these diagrams, and quantitative predictions for them vary greatly between different approaches. In the QCD factorization~(QCDF) approach~\cite{QCDF}, they can only be estimated in a model dependent way due to the endpoint singularities~\cite{Beneke:2001ev}. In the soft-collinear effective theory~(SCET)~\cite{Bauer:2000yr}, they are argued to be factorizable and almost real with tiny strong phase~\cite{Arnesen:2006vb}, which is rather different from almost imaginary with large strong phase as predicted in the perturbative QCD~(pQCD) approach~\cite{Keum:2000ph}. In addition, the annihilation contributions in many $B_{u,d,s}$ decays usually involve both tree and penguin operators, and they interfere with many other different topologies, making it difficult to obtain an accurate value of annihilation by fitting the experimental data~\cite{Du:2002cf}.

 The charmless $B_c$ decays into two light pseudoscalar~(P) and/or vector~(V) mesons, coming only from a single tree operator, provide therefore an ideal testing ground for annihilation in heavy meson decays, and deserve detailed studies using different theoretical approaches~\cite{DescotesGenon:2009ja,Liu:2009qa,Yang:2010ba,Ju:2015SCET}. In this paper, we shall revisit these decays in the QCDF framework, using two different schemes proposed to deal with the endpoint singularity and to avoid enhancements in the soft endpoint region: the divergence in scheme~I is usually parameterized with at least two phenomenological parameters through the treatment $\int^{1}_{0}dx/x \to X_{A,H}={\rm ln}(m_B/\Lambda_h)\,(1+\rho_{A,H} e^{i\phi_{A,H}})$~\cite{Beneke:2001ev}; whereas in scheme~II, one could use an infrared-finite gluon propagator $1/(k^2+i\epsilon)\to1/(k^2-M_g(k^2)+i\epsilon)$~\cite{Cornwall:1981zr}, to regulate the divergent integrals~\cite{BarShalom:2002sv,SuFang,Chang:2008tf,Chang:2012xv,ChangQin}. The different scenarios corresponding to different choices of $\rho_{A,H}$ and $\phi_{A,H}$ in scheme I have been thoroughly discussed in Refs.~\cite{Beneke:2001ev}. In scheme~II, it is found that the hard spectator-scattering contributions are real and the annihilation corrections are complex with a large imaginary part~\cite{Chang:2008tf,Chang:2012xv,ChangQin}. These two different treatments used in $B_{u,d,s}$ decays could be further tested through the charmless $B_c$ decays.

 The remaining parts of the paper are organized as follows. In Sec.~\ref{sec2}, after recapitulating the theoretical framework for two-body charmless hadronic $B_c$ decays, we present the calculation of the annihilation diagrams in the QCDF framework with the two different schemes. The numerical results and discussions are given in Sec.~\ref{sec3}. Finally, we conclude in Sec.~\ref{sec4}. The explicit expressions for the decay amplitudes and the relevant input parameters are collected in Appendix~A and Appendix~B, respectively.

 \section{Theoretical framework and calculation}
 \label{sec2}

 \subsection{The effective weak Hamiltonian and hadronic matrix element}
 \label{sec2.1}

 Using the operator product expansion and renormalization group~(RG) equation, we can write the effective weak Hamiltonian for charmless $B_{c}^{-}\to M_{1}M_{2}$~($M_i$ denote the light pseudoscalar and vector mesons) decays as~\cite{Buras:1998raa}
 \begin{equation}
 {\cal H}_{\rm eff}=\frac{G_{F}}{\sqrt{2}}\,V_{cb}V_{uD}^{\ast}\,\Big[C_{1}(\mu)Q_{1} + C_{2}(\mu)Q_{2}\Big] + {\rm h.c.},
 \label{eq:Heff}
 \end{equation}
where $G_{F}$ is the Fermi coupling constant, $V_{cb}$ and $V_{uD}$~($D=d,s$) the Cabibbo-Kobayashi-Maskawa~(CKM) matrix elements~\cite{CKM}. The four-quark operators $Q_{i}$ arise from $W$-boson exchange and are defined, respectively, as
\begin{eqnarray}
 Q_{1} &= \left[\bar{c}_{\alpha}\gamma^{\mu}(1-\gamma_{5})b_{\alpha}\right] \left[\bar{D}_{\beta}\gamma_{\mu}(1-\gamma_{5})u_{\beta}\right]\,, \nonumber \\
 Q_{2} &= \left[\bar{c}_{\alpha}\gamma^{\mu}(1-\gamma_{5})b_{\beta}\right] \left[\bar{D}_{\beta}\gamma_{\mu}(1-\gamma_{5})u_{\alpha}\right]\,,
 \label{eq:Qi}
\end{eqnarray}
where $\alpha$, $\beta$ are the color indices. The corresponding Wilson coefficients $C_{i}(\mu)$ can be calculated using the RG improved perturbative theory~\cite{Buras:1998raa}.

\begin{figure}[t]
\includegraphics[width=12.0cm]{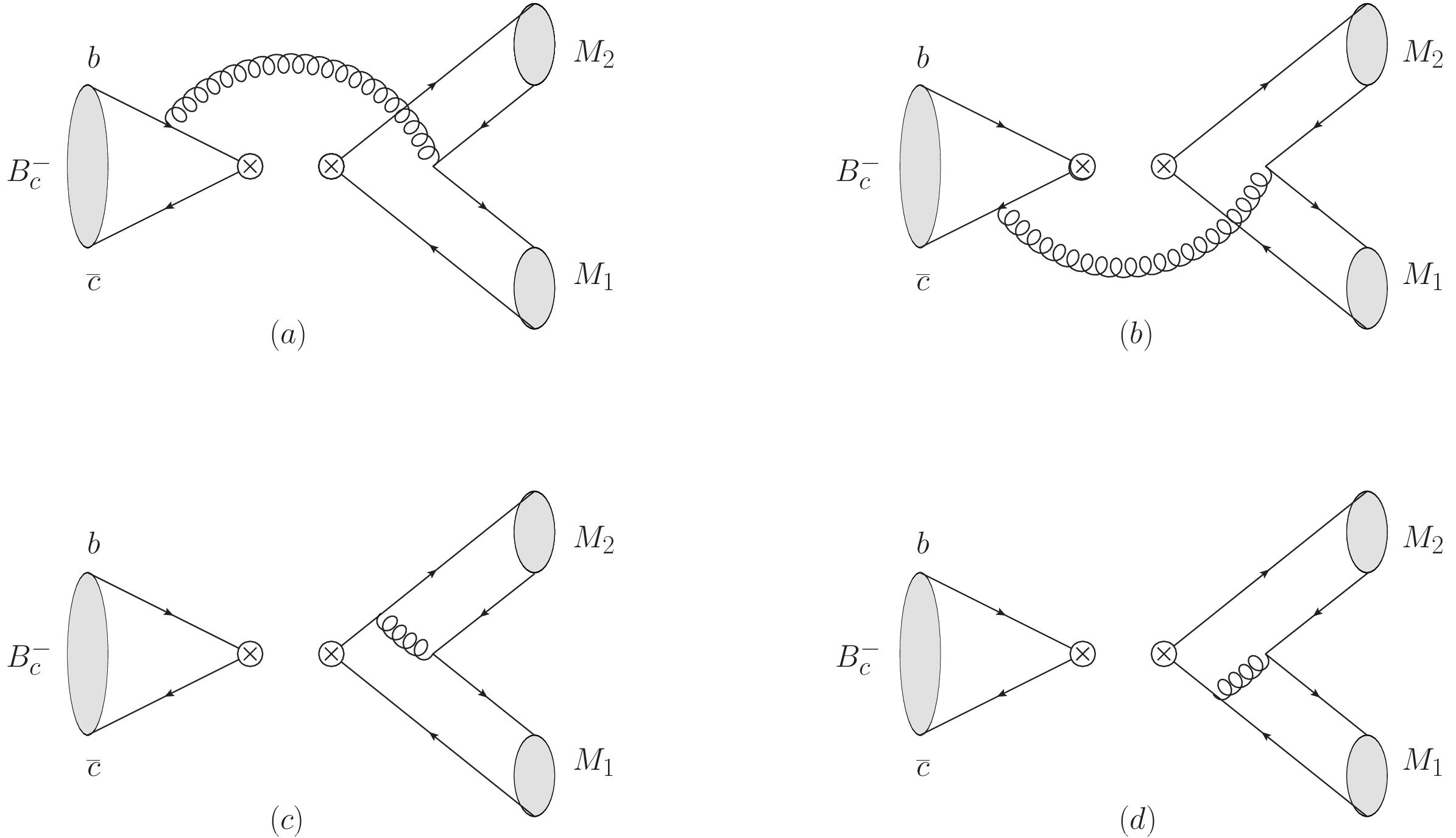}
\caption{The lowest order Feynman diagrams contributing to charmless $B_{c}^{-}\to M_{1}M_{2}$ decays.}
\label{fig.1}
\end{figure}

 To obtain the decay amplitude, the remaining work is to evaluate the hadronic matrix elements of the local operators $Q_i$, which is however quite difficult due to the participation of non-perturbative QCD effects. The Feynman diagrams for $B_{c}\to M_{1}M_{2}$ decays with the QCDF approach are shown in Fig.~\ref{fig.1}, where (a), (b) and (c), (d) are nonfactorizable and factorizable topologies, respectively. Since the tree operators $Q_{1,2}$ have the $(V-A)\otimes(V-A)$ Dirac structure, the two factorizable diagrams (c) and (d) cancel each other exactly in the QCDF approach~\cite{Beneke:2001ev}. Moreover, due to the mismatch of the color indices, there are no contributions from diagrams (a) and (b) with the insertion of the color-singlet operator $Q_1$. Thus, there is only a single tree operator $Q_{2}$ involved in the decay amplitudes, and the nonzero contribution comes only from diagrams (a) and (b).

In the QCDF framework and with the same hypotheses made for hadronic $B_{u,d,s}$ decays, the decay amplitude for charmless $B_{c}^{-}\to M_{1}M_{2}$ decays can be written as~\cite{Beneke:2001ev}
\begin{equation}
 \langle M_{1}M_{2}|{\cal H}_{\rm eff}|B_{c}^{-}\rangle \propto f_{B_{c}}f_{M_{1}}f_{M_{2}}\,b_{2}(M_{1},M_{2})\,,
 \label{eq:matrixelement}
\end{equation}
where $f_{B_{c}}$, $f_{M}$ are decay constants of the $B_{c}$ and $M$ mesons respectively. The coefficient $b_{2}(M_{1},M_{2})$ is defined as~\cite{Beneke:2001ev}
\begin{equation}
b_{2}(M_{1},M_{2})=\frac{C_{F}}{N_{c}^{2}}\,C_{2}\,A_{1}^{i}(M_{1},M_{2})\,,
\end{equation}
where $C_{F}=4/3$ and $N_{c}=3$, the superscript `$i$' on $A_{1}^{i}$ refers to the gluon emission from the initial-state quarks, and the subscript `$1$' on $A_{1}^{i}$ refers to the $(V-A)\otimes(V-A)$ Dirac structure of the inserted four-quark operator $Q_2$. The basic building block $A_{1}^{i}(M_{1},M_{2})$ can be expressed as the convolution of the hard kernels given by diagrams (a) and (b) in Fig.~\ref{fig.1} and the light-cone distribution amplitudes~(LCDAs) of the initial- and final-state mesons, which will be detailed in the next two subsections.

\subsection{$A_{1}^{i}(M_{1},M_{2})$ in scheme~I}
\label{sec2.2}

In scheme~I, the annihilation contributions to hadronic $B_{u,d,s}$ decays are evaluated by regularizing the divergent integrals on the basis of heavy-quark power counting~\cite{Beneke:2001ev}. Despite the fact that such a treatment is not entirely self-consistent in the context of a hard-scattering approach, it provides nevertheless a model to estimate the importance of annihilation, which, motivated by the first observation of the pure annihilation decay $B_s\to \pi^+\pi^-$~\cite{Aaltonen:2011jv,Aaij:2012as}, has been revisited quite recently in Refs.~\cite{Chang:2012xv,Wang:2013fya}.

Following a similar treatment, we now estimate the annihilation topologies in charmless $B_{c}^{-}\to M_{1}M_{2}$ decays. In accordance with the convention adopted in Ref.~\cite{Beneke:2001ev}, we find that the basic building block $A_{1}^{i}(M_{1},M_{2})$ is given by
\begin{eqnarray}
A_{1}^{i}(M_{1},M_{2}) = \pi\alpha_{s}\int_0^1\! dx dy dz \, \Phi_{M_{B_{c}}}(z)\, \Bigg\{\Phi_{M_{2}}(x)\,\Phi_{M_{1}}(y)\,\bigg[\frac{\bar{x}-\bar{z}+z_{b}}{\bar{x}y\big[(\bar{x}+y)\bar{z}-\bar{x}y-i\epsilon\big]} \nonumber\\[0.1cm]
- \frac{y-z+z_{c}}{\bar{x}y\big[(\bar{x}+y)z-\bar{x}y-i\epsilon\big]}\bigg] \nonumber \\[0.2cm]
 + r_{\chi}^{M_{1}}\,r_{\chi}^{M_{2}}\,\Phi_{m_{2}}(x)\,\Phi_{m_{1}}(y)\, \bigg[\frac{\bar{x}yz-x\bar{y}\bar{z}+z_{b}}{\bar{x}y\big[(\bar{x}+y)\bar{z}-\bar{x}y-i\epsilon\big]}
 \nonumber\\[0.1cm]
 - \frac{\bar{x}y\bar{z}-x\bar{y}z+z_{c}}{\bar{x}y\big[(\bar{x}+y)z-\bar{x}y-i\epsilon\big]}\bigg]\Bigg\}\,,
 \label{eq:a1i_schemeI}
\end{eqnarray}
when both mesons are pseudoscalar or when $M_{1}$ is a pseudoscalar and $M_{2}$ a vector meson. In the case when $M_{1}$ is a vector meson and $M_{2}$ a pseudoscalar, one has to change the sign of the second term in $A_{1}^{i}$. When we take $\overline{z}=z_b=1$ and $z=z_c=0$, this result is in agreement with the expressions obtained in Ref.~\cite{Beneke:2001ev,Beneke:2009eb}.
In Eq.~(\ref{eq:a1i_schemeI}), $z_{b}$ and $z_{c}$ denote the relative size of the $b$- and $c$-quark masses with
\begin{equation}
 z_{b}=\frac{m_{b}}{m_{B_c}}\,, \qquad \qquad z_{c}=\frac{m_{c}}{m_{B_c}}\,,
\end{equation}
Their appearance allows one to distinguish the origin of each term in the brackets: the ones involving $z_b$ must come from diagram (a), whereas those involving $z_c$ must from diagram (b) in Fig.~\ref{fig.1}. As always, $\Phi_{M}(x)$ and $\Phi_{m}(x)$ denote the leading-twist and twist-3 two-particle LCDAs of the final-state meson $M$, respectively. The factor $r_{\chi}^{M}$, once multiplied by $m_{B_c}/2$, is used to normalize the twist-3 distribution amplitude; explicitly, we have
\begin{equation}
 r_{\chi}^P(\mu) = \frac{2 m_{P}^2}{m_{B_c}\,[m_1(\mu)+m_2(\mu)]}\,, \qquad \qquad r_\chi^V(\mu) = \frac{2 m_{V}}{m_{B_c}}\,\frac{f_{V}^{\perp}(\mu)}{f_V}\,,
\end{equation}
where $m_{1,2}(\mu)$ denote the running masses of the two valence quarks of a pseudoscalar, and $f_{V}^{\perp}(\mu)$ the scale-dependent transverse decay constant of a vector meson. Despite being formally suppressed by one power of $\mathrm{\Lambda_{QCD}}/m_b$ in the heavy-quark limit, these terms are not always small numerically, especially in the case of pseudoscalar mesons~\cite{Beneke:2001ev}.

In the calculation, we will use three different types of distribution function for $B_{c}$ meson. The first one is the peak form (W-I)~\cite{Brodsky:1985cr}
\begin{equation}
\Phi_{B_c}(x)= \delta(x-m_c/m_{B_c}),
\label{eq:waveI}
\end{equation}
The second one is the solution of the $Schr\ddot{o}dinger$ equation with the harmonic oscillator potential (W-II)~\cite{Sun:2014ika}
\begin{equation}
\Phi_{B_{c}}(x)=Nx\overline{x}\exp\{{-\frac{1}{8\alpha^2}(\frac{m_c^2}{x}+\frac{m_b^2}{\overline{x}})}\},
\label{eq:waveII}
\end{equation}
where $\alpha^2=\mu\omega$, the reduced mass $\mu=m_bm_c/(m_b+m_c)$ and the quantum of energy $\omega\approx0.50$ GeV~\cite{Ebert:2011jc}.
The third one is the quarkonium form (W-III)~\cite{Eilam:2001kw}
\begin{equation}
\Phi_{B_{c}}(x)=Nx\overline{x}\exp\{-(\frac{M_{B_{c}}}{M_{B_{c}}-m_b-m_c})^2
     (x-x_{B_{c}})^2\},
     \label{eq:waveIII}
\end{equation}
where $x_{B_{c}}=1-m_b/m_{B_{c}}$.

\begin{figure}[t]
\includegraphics[width=10.0cm]{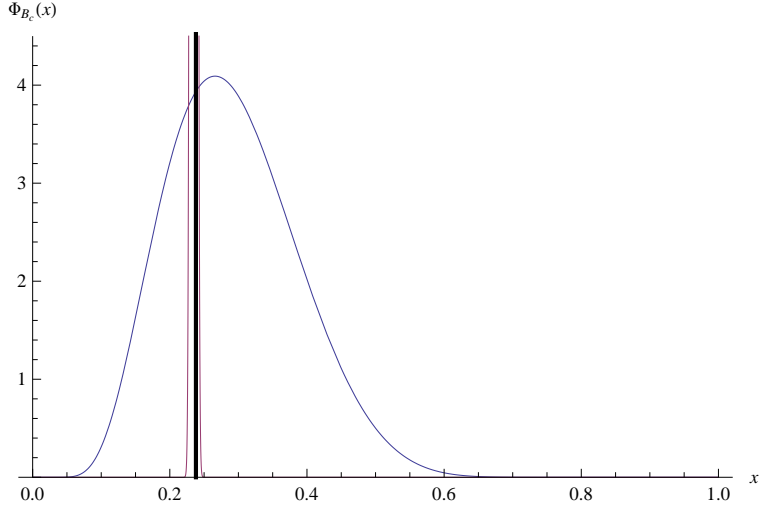}
\caption{The three kinds of distribution function for $B_{c}$ meson, W-I thick solid line,W-II blue line,W-III red line.}
\label{fig.2}
\end{figure}

In Eq.~(\ref{eq:waveII}-\ref{eq:waveIII}), $N$ is normalization constant and the normalization condition is
\begin{equation}
\int_0^1\! dx\Phi_{B_{c}}(x)=1.
\end{equation}
The shape of the three distribution function of $B_{c}$ meson is displayed in Fig.~\ref{fig.2}.

To pursue the structure of the singularities of the building block $A_{1}^{i}(M_{1},M_{2})$, we take, for simplicity, the asymptotic expressions for the distribution amplitudes~\cite{Beneke:2001ev,Ball:1998sk}
\begin{eqnarray}
 &\Phi_{P}(x)=6x(1-x)\,,\qquad \quad \Phi_{V}(x)=6x(1-x)\,,\\
 &\Phi_{p}(x)=1\,,\qquad\qquad \qquad\quad \Phi_{v}(x)=3(2x-1)\,.
\end{eqnarray}

\begin{figure}[t]
\includegraphics[width=10.0cm]{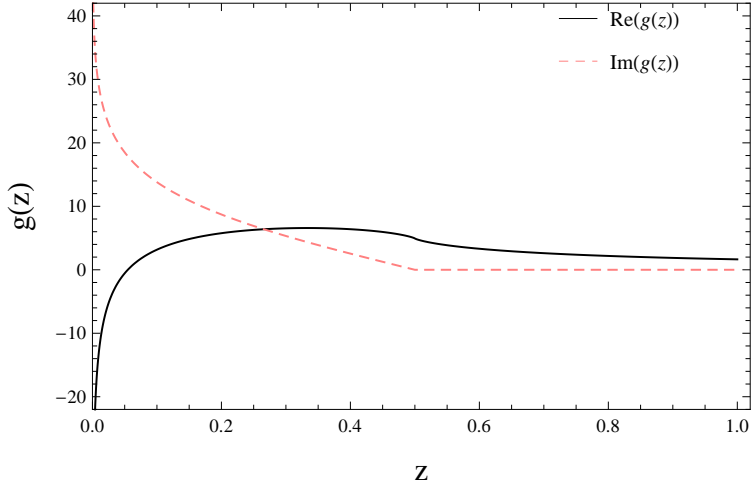}
\caption{The variation of the real and imaginary parts of function $g(z)$ defined by Eq.~(\ref{eq:gz}) with respect to the parameter $z$.}
\label{fig.3}
\end{figure}
The weak annihilation of $B_{u,d,s}$ exhibit endpoint singularities even at twist-2 order in the light-cone expansion for the final-state mesons. For $B_{c}\to PP, PV$ decay, the endpoint singularities only at twist-3 level, has the same situation with the hard spectator interactions of $B_{u,d,s}$. For the twist-2 terms, the singularities in the integral interval. It is found that the convolution integrals in Eq.~(\ref{eq:a1i_schemeI}) can be performed without problem as long as $1/2\leq z<1$. There are, however, integrable singularities at $\bar{x}=z/(1-z)$ or $y=z/(1-z)$ when $0<z<1/2$, which can be dealt with using the prescription of Cauchy principal value integral. Taking the integral
\begin{equation}
 g(z)=\int_0^1\! dx dy\, \frac{1}{(\bar{x}+y)z-\bar{x}y-i\epsilon}
 \label{eq:gz}
\end{equation}
as an example, we show in Fig.~\ref{fig.3} its real and imaginary parts dependence on the parameter $z$, and one can see clearly that the integral is finite as long as $z$ is different from $0$ and $1$. The twist-3 terms in Eq.~(\ref{eq:a1i_schemeI}) is more complex and can not be expressed as polynomial of $X_{A}=\int_0^1{dy}/{y}\sim {\rm ln}(m_b/\Lambda_{QCD})$, so we make the integral interval of $\overline{x},y \in [ \Lambda_{QCD}/m_b, 1 ]$.

Rather than giving the explicit expressions for the convolution integrals, we present, with the default inputs $m_b=4.8~{\rm GeV}$ and $m_c=1.5~{\rm GeV}$, the numerical results for the building block $A_{1}^{i}(M_{1},M_{2})$ in the three different cases \\
W-I
\begin{eqnarray}
 A_{1}^{i}(P,P) &= \pi \left[(-5.70 + 6.26i) + r_{\chi}^{M_{1}}r_{\chi}^{M_{2}}(-1.59 - 2.75i)\right],
  \\[0.2cm]
 A_{1}^{i}(P,V) &= \pi \left[(-5.70 + 6.26i) + r_{\chi}^{M_{1}}r_{\chi}^{M_{2}}(-3.73 + 0.56i)\right],\\[0.2cm]
 A_{1}^{i}(V,P) &= \pi \left[(-5.70 + 6.26i) - r_{\chi}^{M_{1}}r_{\chi}^{M_{2}}( 3.73 - 0.56i)\right],
 \label{eq:a1i_schemeI_num1}
\end{eqnarray}
W-II
\begin{eqnarray}
A_{1}^{i}(P,P)  &= \pi \left[(-2.78 - 4.95i) + r_{\chi}^{M_{1}}r_{\chi}^{M_{2}}(-0.09 - 2.47i)\right],\\[0.2cm]
 A_{1}^{i}(P,V) &= \pi \left[(-2.78 - 4.95i) + r_{\chi}^{M_{1}}r_{\chi}^{M_{2}}(-1.00 - 0.04i)\right],\\[0.2cm]
 A_{1}^{i}(V,P) &= \pi \left[(-2.78 - 4.95i) - r_{\chi}^{M_{1}}r_{\chi}^{M_{2}}( 1.00 + 0.04i)\right],
 \label{eq:a1i_schemeI_num2}
\end{eqnarray}
W-III 
\begin{eqnarray}
 A_{1}^{i}(P,P) &= \pi \left[(-5.82 -6.48i) + r_{\chi}^{M_{1}}r_{\chi}^{M_{2}}(-1.71-2.85i)\right],\\[0.2cm]
 A_{1}^{i}(P,V) &= \pi \left[(-5.82 -6.48i) + r_{\chi}^{M_{1}}r_{\chi}^{M_{2}}(-4.03+0.49i)\right],\\[0.2cm]
 A_{1}^{i}(P,V) &= \pi \left[(-5.82 -6.48i) - r_{\chi}^{M_{1}}r_{\chi}^{M_{2}}( 4.03-0.49i)\right],
 \label{eq:a1i_schemeI_num3}
\end{eqnarray}
where the result is obtained with $\alpha_{s}(\mu\simeq m_{B_c}/2) = 0.25$ and $m_{B_c}=6.2745~{\rm GeV}$. Judging from the above expressions, the branching ratios obtained with W-I and W-III should be very close, and the W-II's results will be smaller. It is noted that $A_{1}^{i}(P,V)$ is identical to $A_{1}^{i}(V,P)$ in our approximation and the annihilation contribution have a large imaginary part.

\subsection{$A_{1}^{i}(M_{1},M_{2})$ in scheme~II}
\label{sec2.3}

Instead of parameterized with an {\it ad hoc} model-dependent cut-off, the endpoint divergences can also be regulated with an infrared~(IR) finite gluon propagator that is characterized by a dynamical gluon mass, providing therefore a natural IR regulator~\cite{Cornwall:1981zr}. This has been successfully applied to various hadronic $B_{u,d,s}$ decays in Refs.~\cite{BarShalom:2002sv,SuFang,Chang:2008tf,Chang:2012xv,ChangQin}. In this subsection, we shall evaluate the building block $A_{1}^{i}(M_{1},M_{2})$ in such a scheme.

Instead of the perturbative expression $1/q^2$ that is IR divergent, the IR finite gluon propagator is obtained by solving an intricate set of coupled Dyson-Schwinger equations~(DSE) for pure gauge QCD, under a systematic approximation and truncation~\cite{Cornwall:1981zr}. It is also noted that any IR finite gluon propagator leads to a freezing of the IR coupling constant~\cite{Aguilar:2002tc}, meaning that the use of an IR finite gluon propagator must be accompanied by an IR finite coupling constant. The above information about the IR behavior of QCD has also been confirmed by the most recent lattice simulations~\cite{Aouane:2012bk,Gongyo:2013sha}\footnotemark[1]\footnotetext[1]{Recent reviews, together with a list of references, on DSE solutions and lattice results about the infrared finite gluon propagator and running coupling constant could be found, for example, in Refs.~\cite{Cornwall:2009ud,Alkofer:2000wg,Fischer:2008uz,Boucaud:2011ug}.}. Here we adopt the gluon propagator derived by Cornwall many years ago~\cite{Cornwall:1981zr}
\begin{equation}
 D(q^{2})=\frac{1}{q^2+M_{g}^2(q^2)}\,,
 \label{eq:gluonprop}
\end{equation}
where $q^2$ denotes the gluon momentum squared. The corresponding running coupling constant reads~\cite{Cornwall:1981zr}
\begin{equation}
 \alpha_{s}(q^2)=\frac{4\pi}{\beta_{0}\ln\left[\frac{q^2+4M_{g}^2(q^2)}{\Lambda_{\rm QCD}^2}\right]}\,,
 \label{eq:alphas}
\end{equation}
where $\beta_{0}=11-\frac{2}{3}n_{f}$ is the first coefficient of the QCD beta function, and $n_{f}$ the number of active quark flavors at a given scale. The dynamical gluon mass $M_{g}^{2}(q^{2})$ is given by~\cite{Cornwall:1981zr}
\begin{equation}
 M_{g}^{2}(q^{2}) = m_{g}^2 \left[\frac{\ln(\frac{q^{2}+4m_{g}^{2}}{\Lambda_{\rm QCD}^{2}})}
{\ln(\frac{4m_{g}^{2}}{\Lambda_{\rm QCD}^{2}})}\right]^{-\frac{12}{11}}\,,
\label{eq:effmass}
\end{equation}
where $\Lambda_{\rm QCD}=225~{\rm MeV}$ is the QCD scale, and $m_{g}$ the effective gluon mass with a typical value $m_{g}=0.5\pm0.2~{\rm GeV}$~\cite{Cornwall:1981zr}. It is interesting to note that similar values are found by fitting the experimental data on $B_{u,d,s}$ decays:  $m_{g}=0.5\pm0.05~{\rm GeV}$ from $B_{u,d}$ decays~\cite{Chang:2008tf} and $m_{g}=0.48\pm0.02~{\rm GeV}$ from $B_s$ decays~\cite{Chang:2012xv}. In our calculation, we take $m_{g}=0.49\pm0.03~{\rm GeV}$. As shown in Fig.~\ref{fig.3}, both the gluon propagator~(Eq.~(\ref{eq:gluonprop})) and the coupling constant~(Eq.~(\ref{eq:alphas})) are IR finite and different from zero at the origin of momentum squred $q^2=0$.

\begin{figure}[t]
\includegraphics[width=7.0cm]{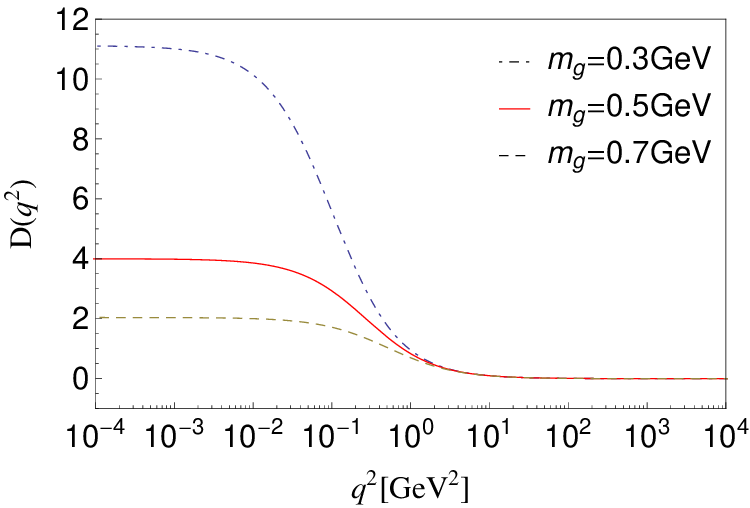}  \qquad
\includegraphics[width=7.0cm]{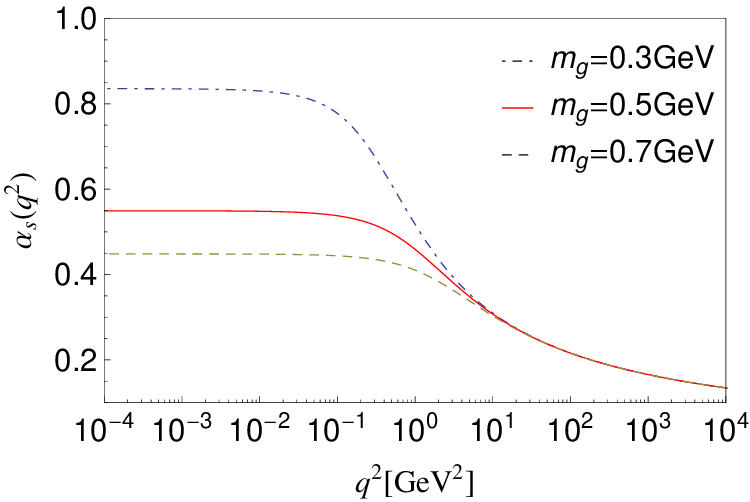}
\caption{The behavior of the gluon propagator~(left) and coupling constant~(right) derived by Cornwall~\cite{Cornwall:1981zr}, with respect to the gluon momentum squred $q^2$.}
\label{fig.4}
\end{figure}

With the above prescription and the same convention used in scheme~I, our final results for the building block $A_{1}^{i}(M_{1},M_{2})$ can be expressed as~($\omega^{2}(q^2)=M_{g}^{2}(q^{2})/m_{B_c}^2$)
\begin{eqnarray}
 A_{1}^{i}(M_{1},M_{2}) &=& \pi \int_{0}^{1}\! dx dy dz \,\alpha_{s}(q^{2})\, \Phi_{M_{B_{c}}}(z)\,\Bigg\{\quad\quad\quad\quad\quad \quad\quad \nonumber\\[0.1cm] & & \quad\quad\quad\quad \Phi_{M_{2}}(x)\,\Phi_{M_{1}}(y)\,\bigg[\frac{\bar{x}-\bar{z}+z_{b}}{(\bar{x}y-\omega^{2}(q^2)+i\epsilon)\,
 \big[(\bar{x}+y)\bar{z}-\bar{x}y-i\epsilon\big]} \nonumber\\[0.1cm]
 & &\quad\quad\quad\quad\quad\quad\quad\quad\quad\quad- \frac{y-z+z_{c}}{(\bar{x}y-\omega^{2}(q^2)+i\epsilon)\,\big[(\bar{x}+y)z-\bar{x}y-i\epsilon\big]}\bigg] \nonumber\\[0.2cm]
& & + r_{\chi}^{M_{1}}\,r_{\chi}^{M_{2}}\,\Phi_{m_2}(x)\,\Phi_{m_1}(y)\,\bigg[\frac{\bar{x}yz-x\bar{y}\bar{z}+z_{b}}
 {(\bar{x}y-\omega^{2}(q^2)+i\epsilon)\,\big[(\bar{x}+y)\bar{z}-\bar{x}y-i\epsilon\big]} \nonumber\\[0.1cm]
 & &\quad\quad\quad\quad\quad\quad\quad\quad\quad\quad- \frac{\bar{x}y\bar{z}-x\bar{y}z+z_{c}}{(\bar{x}y-\omega^{2}(q^2)+i\epsilon)\,\big[(\bar{x}+y)z-\bar{x}y-i\epsilon\big]}\bigg]\Bigg\}\,,
 \label{eq:a1i_schemeII}
\end{eqnarray}
when $(M_{1},M_{2})=(P,P)$ and $(P,V)$. If $(M_{1},M_{2})=(V,P)$, on the other hand, the sign of the second term in $A_{1}^{i}$ has to be changed. Our results, after taking the limits $\bar{z}=z_b\to 1$ and $z=z_c\to 0$, agree with the ones given in Refs.~\cite{Chang:2008tf,Chang:2012xv,ChangQin}.

\begin{figure}[t]
\includegraphics[width=8.0cm]{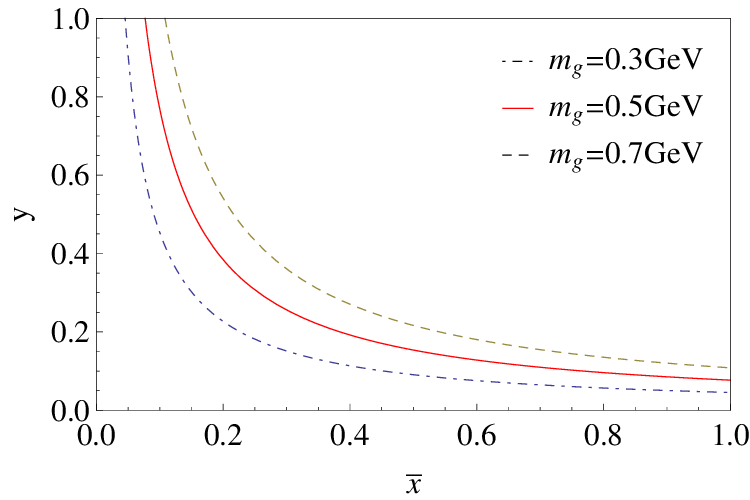}
\caption{The distribution in the $(\bar{x},y)$ plane of the IR finite gluon propagator appearing in the annihilation diagrams for charmless hadronic $B_c\to M_1 M_2$ decays.}
\label{fig.5}
\end{figure}

In Eq.~(\ref{eq:a1i_schemeII}), the time-like gluon momentum squared is given by $q^2=\bar{x}ym_{B_c}^2$, and also depends on the longitudinal momentum fractions $\bar{x}=1-x$ and $y$, making the convolution integrals rather complicated. As shown in Fig.~\ref{fig.5}, although the running coupling constant is rather large in the small $q^2$ region~(see Fig.~\ref{fig.4}(b)), the fact that only a small fraction comes from the $q^2<m_g^2$ regions in the $(\bar{x},y)$ plane indicates that the annihilation contributions are still dominated by the $q^2>m_g^2$ regions associated with a large imaginary part~\cite{Chang:2008tf}. In scheme II, we also use three kinds of $B_{c}$ meson wave function in the calculation. With the default inputs $m_{B_c}=6.2745~{\rm GeV}$, $m_{g}=0.49~{\rm GeV}$, $m_b=4.8~{\rm GeV}$ and $m_c=1.5~{\rm GeV}$, our numerical results for the building block $A_{1}^{i}(M_{1},M_{2})$ read\\
W-I
\begin{eqnarray}
A_{1}^{i}(P,P)&=\pi \left[(-10.11 - 6.16i) + r_{\chi}^{M_{1}}r_{\chi}^{M_{2}}(-3.64 - 2.36i)\right],\\[0.2cm]
A_{1}^{i}(P,V)&=\pi \left[(-10.11 - 6.16i) + r_{\chi}^{M_{1}}r_{\chi}^{M_{2}}(-6.38 + 3.09i)\right],\\[0.2cm]
A_{1}^{i}(V,P)&=\pi \left[(-10.11 - 6.16i) - r_{\chi}^{M_{1}}r_{\chi}^{M_{2}}(6.38 - 3.09i)\right],
\label{eq:a1i_schemeII_num1}
\end{eqnarray}
W-II 
\begin{eqnarray}
A_{1}^{i}(P,P)&=\pi \left[(-5.84 - 6.06i) + r_{\chi}^{M_{1}}r_{\chi}^{M_{2}}(-4.24 - 7.76i)\right],\\[0.2cm]
A_{1}^{i}(P,V)&=\pi \left[(-5.84 - 6.06i) + r_{\chi}^{M_{1}}r_{\chi}^{M_{2}}(-18.36 - 9.27i)\right],\\[0.2cm]
A_{1}^{i}(V,P)&=\pi \left[(-5.84 - 6.06i) - r_{\chi}^{M_{1}}r_{\chi}^{M_{2}}(12.5 + 9.94i)\right],
\label{eq:a1i_schemeII_num2}
\end{eqnarray}
W-III 
\begin{eqnarray}
A_{1}^{i}(P,P)&=\pi \left[(-10.37 - 6.36i) + r_{\chi}^{M_{1}}r_{\chi}^{M_{2}}(-3.57 - 1.89i)\right],\\[0.2cm]
A_{1}^{i}(P,V)&=\pi \left[(-10.37 - 6.36i) + r_{\chi}^{M_{1}}r_{\chi}^{M_{2}}(-5.97 + 4.29i)\right],\\[0.2cm]
A_{1}^{i}(V,P)&=\pi \left[(-10.37 - 6.36i) - r_{\chi}^{M_{1}}r_{\chi}^{M_{2}}(6.02 - 4.30i)\right].
\label{eq:a1i_schemeII_num3}
\end{eqnarray}
One can see that, compared to the values obtained in scheme~I~(Eq.~(\ref{eq:a1i_schemeI_num1})-(\ref{eq:a1i_schemeI_num3})),  the annihilation contributions predicted in scheme~II are enhanced. This will apparently affect the predictions for charmless hadronic $B_c\to M_1M_2$ decays, which will be detailed in the next section.

\section{Numerical results and discussions}
\label{sec3}

In the $B_c$-meson rest frame, the branching ratio for a general charmless $B_{c}^{-}\to M_1M_2$ decay can be written as\footnote{At the moment we focus only on the $PP$, $PV$, and $VP$ modes. For the $VV$ mode, three different configurations for the outgoing mesons, labeled by their helicities, have to be considered. In the QCDF approach, the transversely polarised amplitudes in $VV$ modes do not factorise even at leading power in the heavy-quark expansion, making the calculation less predictive~\cite{Beneke:2001ev,Beneke:2009eb}.}
\begin{equation}
 {\rm Br}(B_{c}^{-}\to M_1M_2) = \frac{\tau_{B_c}}{8\pi}\,\frac{|\vec{p}|}{m_{B_c}^2} \big|{\cal A}(B_{c}^{-}\to M_1M_2)\big|^2\,,
\end{equation}
where $\tau_{B_c}$ is the $B_c$-meson lifetime, and $|\vec{p}|$ is the center-of-mass momentum of either of the two outgoing mesons, with
\begin{equation}
|\vec{p}|=\frac{\sqrt{\left[m_{B_c}^2-(m_{M_1}+m_{M_2})^2\right]\left[m_{B_c}^2-(m_{M_1}-m_{M_2})^2\right]}}{2 m_{B_c}}\,.
\end{equation}
The decay amplitude ${\cal A}(B_{c}^{-}\to M_1M_2)$ can be obtained from the hadronic matrix element $\langle M_{1}M_{2}|{\cal H}_{\rm eff}|B_{c}^{-}\rangle$ defined in Eq.~(\ref{eq:matrixelement}); for convenience, we collect in Appendix~A the explicit expressions of the decay amplitudes for the considered decay modes. The CP-violating asymmetries for all the considered $B_c$ decays are absent, since there involves only a single tree operator in the decay amplitudes, which can be clearly seen from Eq.~(\ref{eq:matrixelement}).

With the theoretical expressions given above and the input parameters collected in Appendix~B, we proceed to evaluate the CP-averaged branching ratios for these charmless $B_{c}\to M_{1}M_{2}$ decays. In our calculation, the default value of the renormalization scale is set at $\mu=m_{B_c}/2$, which is approximately the averaged virtuality of the time-like gluon propagated in the annihilation diagrams. The numerical results based on the two schemes are collected in Table~\ref{tab01}. In Table \ref{tab02}, Table \ref{tab03} and Table \ref{tab04}, we also present detailed error estimates induced by the theoretical uncertainties of input parameters for the strangeness-conserving~($|\triangle S|=0$) processes. The first error shown corresponds to the variation of the CKM parameters $A$ and $\lambda$~(named as ``CKM"), the second error refers to the variation of the quark masses, decay constants, and the $\eta-\eta^{\prime}$ mixing angle~(named as ``hadronic"). The third error arises from the variation of the renormalization scale $\mu$~(named as ``scale"). The last error reflects the uncertainty due to the dynamical gluon mass $m_g$~(named as ``$m_g$").

Based on the results collected in Table~\ref{tab01}-\ref{tab04}, we have the following observations and remarks:
\begin{itemize}
 \item The two-body charmless hadronic $B_{c}\to M_{1}M_{2}$ decays can be classified into two categories: the strangeness-conserving~($|\triangle S|=0$) and the strangeness-changing~($|\triangle S|=1$) processes. From the numerical results listed in Table~\ref{tab01}, one can see that the branching rations of $|\triangle S|=0$ channels are generally much larger than those of $|\triangle S|=1$ ones. This is due to the large hierarchical structure between the two CKM matrix elements $V_{ud}$ and $V_{us}$, $|V_{ud}/V_{us}|^2\sim 19$.

 \item In scheme I, the branching ratios obtained with W-I  and W-III very close, they are vary in the ranges of $10^{-10}$ to $10^{-8}$, being larger than the corresponding ones obtained with W-II. This is consistent with the wave function, W-III is very close to $\delta$ function as shown in Fig.~\ref{fig.2}. In scheme II, the annihilation contribution are enhanced when we adopt the IR finite gluon propagator, and the branching ratios are not sensitive to the choice of wave function for $B_{c}$ meson. On the whole, the results of this paper are smaller than the corresponding ones obtained in the pQCD approach~\cite{Liu:2009qa}. The large discrepancies among these theoretical predictions make it very necessary to make more detailed studies of these kinds of $B_c$ decays.

 \item Among these charmless $B_{c}$ decays, only several decays modes, such as $B_{c}^{-}\to K^{-}K^{0}$, $K^{*-}K^{0}$, $K^{-}K^{*0}$, $\pi^{-}\omega$, and $\rho^{-}\eta^{(')}$, have relatively large branching ratios, being around ${\cal O}(10^{-7})$ in Scheme-II. All of these channels belong to the $|\triangle S|=0$ transitions that are CKM favored. It is found that branching ratio for $B_{c}^{-}\to \pi^{-}\omega$ decay has the largest branching ratio ${\rm Br}(B_{c}^{-}\to \pi^{-}\omega)=12.8\times10^{-8}$, which is promisingly detected by experiments at the running Large Hadron Collider and forthcoming SuperKEKB.

 \item For $B_{c}^{-}\to\pi^{-}\pi^{0}$, $\pi^{-}\rho^{0}$, and $\rho^{-}\pi^{0}$ decays, on the other hand, since $|\pi^0\rangle,|\rho^0\rangle=(|\bar{u}u\rangle-|\bar{d}d\rangle)/\sqrt{2}$, the contributions from $\bar{u}u$ and $\bar{d}d$ components of the neutral mesons cancel each other exactly or almostly, resulting in (approximate) zero branching ratios of these three channels. For $B_{c}^{-}\to \pi^{-}\omega$, $\pi^{-}\eta^{(\prime)}$ and $\rho^{-}\eta^{(\prime)}$ decays, due to the flavor decomposition $|\omega\rangle, |\eta_q\rangle=(|\bar{u}u\rangle+|\bar{d}d\,\rangle)/\sqrt{2}$, the interference between the two flavor components $\bar{u}u$ and $\bar{d}d$ of the neutral mesons is constructive, resulting in larger branching ratios. Taking into account the fact that $f_{\omega}>f_{\eta}>f_{\eta^{\prime}}$, one can easily understand the pattern of their branching ratios. Especially, the decay modes $\pi^{-}(\rho^{-})\eta$ and $\pi^{-}(\rho^{-})\eta^{\prime}$ have similar branching ratios, because only the $|\eta_q\rangle$ term involves in the decay amplitudes.

 \item For $B_{c}^{-}\to K^{(*)-}\eta^{(\prime)}$ decays, the obtained branching ratios show a rather different pattern, ${\rm Br}(K^{(*)-}\eta^{\prime})\gg {\rm Br}(K^{(*)-}\eta)$, from that of ${\rm Br}(\pi^{-}(\rho^{-})\eta^{\prime})\sim{\rm Br}(\pi^{-}(\rho^{-})\eta)$. It is also observed that ${\rm Br}(K^{(*)-}\eta^{\prime})$ is much larger than ${\rm Br}(K^{(*)-}\pi^0)$, while ${\rm Br}(K^{(*)-}\eta)$ is suppressed rather than enhanced compared to that of the $\pi^{0}$ mode. To understand the enhancement and suppression patterns, we should note that both the $|\eta_q\rangle$ and $|\eta_s\rangle$ terms contribute to these $|\triangle S|=1$ transitions, but with an opposite sign between them for the $\eta$ and $\eta^{\prime}$ final states, which is due to the fact that $f_{\eta^{(\prime)}}^q>0$, $f_{\eta^{\prime}}^s>0$, while $f_{\eta}^s<0$. This results in a destructive interference for the $\eta$, but a constructive interference for the $\eta^{\prime}$ modes. Similar patterns have already been observed in the $B\to K\eta^{(\prime)}$ and $B\to K^{*}\eta^{(\prime)}$ decays~\cite{Beneke:2002jn}.

 \item As discussed in Ref.~\cite{DescotesGenon:2009ja}, several relations among the charmless $B_c$ decay channels hold in the limit of exact SU(3) flavor symmetry. For $B_{c}\to PP$ decays, for example, one of such relations reads
     \begin{equation}
     {\cal A}(B_{c}^{-}\to \bar{K}^{0}\pi^{-})=\sqrt{2}{\cal A}(B_{c}^{-}\to K^{-}\pi^{0})=\hat{\lambda}{\cal A}(B_{c}^{-}\to K^{-}K^{0})\,,
     \end{equation}
     with the Cabibbo-suppressing factor $\hat{\lambda}=V_{us}/V_{ud}$. Similar relations could also be found for $B_{c}\to PV$ decays, with the replacements $\pi\to\rho$ and/or $K\to K^{*}$. We find that the first equality holds exactly in both scheme I and scheme II, because the exact isospin symmetry is assumed in our calculation. The second equality is, however, violated by the differences between decay constants and light-quark masses, which account for the SU(3)-breaking effect.
 \item As the relevant CKM parameters have been measured quite precisely, the theoretical errors introduced by the CKM parameters are small. The uncertainty due to the variation of the dynamical gluon mass in scheme II is also found to be negligible. The main uncertainties are due to the variation of the renormalization scale, as well as the mixing parameters for $\eta$ and $\eta^{\prime}$ final states. The large scale dependence of the branching ratios is understandable, because only the leading order term in $\alpha_s$ is taken into account in our calculation. Furthermore, the different choices of the renormalization scale also account for the main differences among our results and the ones presented in Refs.~\cite{DescotesGenon:2009ja,Liu:2009qa}.
\end{itemize}

Finally, we would like to point out that it is hard to estimate the systematical uncertainties coming from the hypothesis underlying our calculations, such as the one-gluon approximation for the annihilation mechanism, the use of asymptotic distribution amplitudes, as well as the neglect of $1/m_b$-suppressed power corrections.

\section{Summary}
\label{sec4}

Being the lowest-lying bound state of two heavy quarks with different flavors, the $B_{c}$ meson is an ideal system to study weak decays of heavy mesons. In this paper, we have carried out a detailed study of two-body charmless hadronic $B_c$ decays, which can proceed only via the weak annihilation diagrams within the SM and are, therefore, very suitable for further improving our understanding of the annihilation mechanism, the size of which is currently an important issue in $B$ physics. Explicitly, we have adopted two different schemes to deal with these decays: scheme I is similar to the usual method adopted in the QCDF approach, while scheme II is based on the infrared behavior of gluon propagator and running coupling. For comparison, we adopt three different kinds of distribution function for $B_{c}$ meson. It is found that the strength of annihilation contributions predicted in scheme II is enhanced compared to that obtained in scheme I. The branching ratios are not sensitive to the choice of wave function for $B_{c}$ meson in scheme II. However, the predicted branching ratios are inconsisitent with the corresponding ones obtained in the pQCD approach~\cite{Liu:2009qa}.

The large discrepancies among these theoretical predictions make it very necessary to make more detailed studies of these kinds of $B_c$ decays, especially from the experimental side. It is interesting to note that the LHCb experiment has the potential to observe the decays with a branching ratio of $10^{-7}$, which will certainly provide substantial information on these charmless $B_{c}$ decays and deepen our understanding of the annihilation mechanisms.

\section*{Acknowledgements}
 This work is supported by CCNU-QLPL Innovation Fund (QLPL201411).

\begin{appendix}

\section{Decay amplitudes in the QCDF approach}
\label{appendix A}

Starting with Eq.~(\ref{eq:matrixelement}) and adopting the standard phase convention for the flavor wave functions of light and heavy mesons~\cite{Beneke:2001ev,Feldmann:1998vh,Beneke:2002jn}, one can easily write down the decay amplitude for a given decay mode. Firstly, there are eight charmless $B_c\to PP$ decays with the corresponding amplitude given, respectively, as~(the exact isospin symmetry is assumed):
\begin{eqnarray}
{\cal A}(B_{c}^{-}\to \pi^{-}\pi^{0}) &= \frac{G_{F}}{2}\,V_{cb}V_{ud}^{*}\,f_{B_{c}}f_{\pi^{-}}
f_{\pi^{0}} \left[b_{2}(\pi^{0}, \pi^{-})-b_{2}(\pi^{-}, \pi^{0})\right]=0\,,\\[0.2cm]
{\cal A}(B_{c}^{-}\to \pi^{-}\eta^{(\prime)}) &= \frac{G_{F}}{2}\,V_{cb}V_{ud}^{*}\,f_{B_{c}}f_{\pi^{-}} f_{\eta^{(\prime)}}^{q} \left[b_{2}(\pi^{-}, \eta^{(\prime)})+b_{2}(\eta^{(\prime)}, \pi^{-})\right]\,,\\[0.2cm]
{\cal A}(B_{c}^{-}\to K^{-} K^{0}) &= \frac{G_{F}}{\sqrt{2}}\,V_{cb}V_{ud}^{*}\,f_{B_{c}}f_{K^{-}}f_{K^{0}}\, b_{2}(K^{-},K^{0})\,, \\[0.2cm]
{\cal A}(B_{c}^{-}\to K^{-}\pi^{0}) &= \frac{G_{F}}{2}\,V_{cb}V_{us}^{*}\,f_{B_{c}}f_{K^{-}}f_{\pi^{0}}\, b_{2}(\pi^{0},K^{-})\,, \\[0.2cm]
{\cal A}(B_{c}^{-}\to \bar{K}^{0}\pi^{-}) &= \frac{G_{F}}{\sqrt{2}}\,V_{cb}V_{us}^{*}\,f_{B_{c}} f_{K^{0}}f_{\pi^{-}}\,b_{2}(\pi^{-},\bar{K}^{0})\,, \\[0.2cm]
{\cal A}(B_{c}^{-}\to K^{-}\eta^{(\prime)}) &= \frac{G_{F}}{2}\,V_{cb}V_{us}^{*}\,f_{B_{c}}
f_{K^{-}}\left[f_{\eta^{(\prime)}}^{q}\,b_{2}(\eta^{(\prime)},K^{-}) + \sqrt{2}f_{\eta^{(\prime)}}^{s}\,b_{2}
(K^{-},\eta^{(\prime)})\right]\,.
\end{eqnarray}
The decay amplitudes for the $15$ charmless $PV$ modes can be written, respectively, as:
\begin{eqnarray}
\mathcal{A}(B_{c}^{-}\to \pi^{-}\rho^{0}) &= \frac{G_{F}}{2}\,V_{cb}V_{ud}^{*}\,f_{B_{c}}f_{\pi^{-}}
f_{\rho^{0}}\left[b_{2}(\rho^{0}, \pi^{-})-b_{2}(\pi^{-}, \rho^{0})\right]\,, \\[0.2cm]
\mathcal{A}(B_{c}^{-}\to \pi^{-}\omega) &= \frac{G_{F}}{2}\,V_{cb}V_{ud}^{*}\,f_{B_{c}}f_{\pi^{-}}
f_{\omega}\left[b_{2}(\omega, \pi^{-})+b_{2}(\pi^{-}, \omega)\right]\,, \\[0.2cm]
\mathcal{A}(B_{c}^{-}\to K^{*-}K^{0}) &= \frac{G_{F}}{\sqrt{2}}\,V_{cb}V_{ud}^{*}\,f_{B_{c}}f_{K^{*-}}f_{K^{0}}\, b_{2}(K^{*-}, K^{0})\,, \\[0.2cm]
\mathcal{A}(B_{c}^{-}\to K^{-}\rho^{0}) &= \frac{G_{F}}{2}\,V_{cb}V_{us}^{*}\,f_{B_{c}}f_{K^{-}}f_{\rho^{0}}\, b_{2}(\rho^{0}, K^{-})\,, \\[0.2cm]
\mathcal{A}(B_{c}^{-}\to \bar{K}^{0}\rho^{-}) &= \frac{G_{F}}{\sqrt{2}}\,V_{cb}V_{us}^{*}\,f_{B_{c}}f_{K^{0}}
f_{\rho^{-}}\,b_{2}(\rho^{-}, \bar{K}^{0})\,, \\[0.2cm]
\mathcal{A}(B_{c}^{-}\to K^{-}\omega) &= \frac{G_{F}}{2}\,V_{cb}V_{us}^{*}\,f_{B_{c}}f_{K^{-}}f_{\omega}\,
b_{2}(\omega, K^{-})\,,\\[0.2cm]
\mathcal{A}(B_{c}^{-}\to \rho^{-}\pi^{0}) &= \frac{G_{F}}{2}\,V_{cb}V_{ud}^{*}\,f_{B_{c}}f_{\rho^{-}}
f_{\pi^{0}}\left[b_{2}(\pi^{0}, \rho^{-})-b_{2}(\rho^{-}, \pi^{0})\right]\,, \\[0.2cm]
\mathcal{A}(B_{c}^{-}\to \rho^{-}\eta^{(\prime)}) &= \frac{G_{F}}{2}\,V_{cb}V_{ud}^{*}\,f_{B_{c}}f_{\rho^{-}}
f_{\eta^{(\prime)}}^{q}\left[b_{2}(\eta^{(\prime)}, \rho^{-})+b_{2}(\rho^{-}, \eta^{(\prime)})\right]\,, \\[0.2cm]
\mathcal{A}(B_{c}^{-}\to K^{-}K^{*0}) &= \frac{G_{F}}{\sqrt{2}}\,V_{cb}V_{ud}^{*}\,f_{B_{c}}f_{K^{-}}f_{K^{*0}}\, b_{2}(K^{-}, K^{*0})\,,\\[0.2cm]
\mathcal{A}(B_{c}^{-}\to K^{*-}\pi^{0}) &= \frac{G_{F}}{2}\,V_{cb}V_{us}^{*}\,f_{B_{c}}f_{K^{*-}}f_{\pi^{0}}\, b_{2}(\pi^{0}, K^{*-})\,, \\[0.2cm]
\mathcal{A}(B_{c}^{-}\to \bar{K}^{*0}\pi^{-}) &= \frac{G_{F}}{\sqrt{2}}\,V_{cb}V_{us}^{*}\,f_{B_{c}}f_{K^{*0}} f_{\pi^{-}}\,b_{2}(\pi^{-}, \bar{K}^{*0})\,, \\[0.2cm]
\mathcal{A}(B_{c}^{-}\to K^{*-}\eta^{(\prime)}) &= \frac{G_{F}}{2}\,V_{cb}V_{us}^{*}\,f_{B_{c}} f_{K^{*-}} \left[f_{\eta^{(\prime)}}^{q}\,b_{2}(\eta^{(\prime)}, K^{*-})+\sqrt{2}f_{\eta^{(\prime)}}^{s}\,b_{2}
(K^{*-}, \eta^{(\prime)})\right]\,, \\[0.2cm]
\mathcal{A}(B_{c}^{-}\to \phi K^{-}) &= \frac{G_{F}}{\sqrt{2}}\,V_{cb}V_{us}^{*}\,f_{B_{c}}f_{\phi}
f_{K^{-}}\,b_{2}(K^{-}, \phi)\,.
\end{eqnarray}

\section{Input parameters}
\label{appendix B}

To get the Wilson coefficients $C_i(\mu)$ at the lower scale $\mu=m_{B_c}/2$, we adopt the following input parameters~\cite{PDG2014}:
\begin{eqnarray}
& \alpha_s(M_Z)=0.1185\pm0.0006, \qquad \alpha(M_Z)=1/128, \qquad \sin^2\theta_W=0.23, \nonumber \\
& M_Z=91.1876~{\rm GeV}, \quad M_W=80.385~{\rm GeV}, \quad m_t=173.21\pm0.87~{\rm GeV}.
\end{eqnarray}
We also vary the renormalization scale $\mu$ in the region $[m_{B_c}/4,m_{B_c}]$ to assess the scale uncertainty.

For the CKM matrix elements, we use the Wolfenstein parameterization~\cite{Wolfenstein:1983yz} and keep terms up to $\mathcal{O}(\lambda^4)$~\cite{Buras:1998raa}:
\begin{eqnarray}
V_{ud}&=1-\frac{1}{2}\lambda^2-\frac{1}{8}\lambda^4+\mathcal{O}(\lambda^6)\,, \nonumber \\
V_{us}&=\lambda+\mathcal{O}(\lambda^7)\,, \qquad V_{cb}=A\lambda^2+\mathcal{O}(\lambda^8)\,,
\end{eqnarray}
with the inputs $A=0.813^{+0.015}_{-0.027}$ and $\lambda=0.22551^{+0.00068}_{-0.00035}$~\cite{Charles:2004jd}.

For the $\eta-\eta^{\prime}$ system, we adopt the Feldmann-Kroll-Stech~(FKS) mixing scheme defined in the quark-flavor basis~\cite{Feldmann:1998vh}, where the physical states $|\eta\rangle$ and $|\eta^{\prime}\rangle$ are related to the flavor states $|\eta_q\rangle=(|\bar{u}u\rangle+|\bar{d}d\,\rangle)/\sqrt{2}$ and $|\eta_s\rangle=|s\bar{s}\rangle$ by
\begin{equation}
\left(\begin{array}{c} |\eta\rangle \\ |\eta^{\prime}\rangle \end{array}\right)
=\left(\begin{array}{lr} \cos\phi & \quad -\sin\phi \\ \sin\phi & \quad \cos\phi \end{array}\right)
\left(\begin{array}{c} |\eta_{q}\rangle \\ |\eta_{s}\rangle \end{array}\right)\,.
\end{equation}
The decay constants $f_{\eta^{(\prime)}}^{q}$ and $f_{\eta^{(\prime)}}^{s}$, as well as the other hadronic parameters related to $\eta$ and $\eta^{\prime}$ can then be expressed in terms of two decay constants $f_{q,s}$ and the mixing angle $\phi$~\cite{Beneke:2002jn}. The values of these three parameters have been determined from a fit to experimental data, yielding~\cite{Feldmann:1998vh}
\begin{equation}
f_{q}=(1.07\pm0.02)f_{\pi}, \qquad  f_{s}=(1.34\pm0.06)f_{\pi}, \qquad \phi=39.3^{\circ}\pm1.0^{\circ}.
\end{equation}

Finally, a summary of the other input parameters entering our numerical analysis is given in Table~\ref{tab:inputs}. It is noted that the latest experimental determinations of $f_{\pi}$ and $f_K$~\cite{PDG2014} compare positively within errors with the lattice results~\cite{Aoki:2013ldr}. Our values of the vector-meson decay constants are taken from Ref.~\cite{Jung:2012vu}, which are an update of the ones extracted in Ref.~\cite{Ball:2006eu}. The scale dependence of the transverse decay constants is taken into account via the leading-logarithmic running $f_\perp(\mu) = f_\perp(\mu_0) \,\left[\alpha_s(\mu)/\alpha_s(\mu_0)\right]^{4/23}$. The light quark masses given in the table are the running masses defined in the ${\rm \overline{MS}}$ scheme; to get the corresponding pole and running masses at different scales, we use the NLO running formulae collected, for example, in Ref.~\cite{Chetyrkin:2000yt}. The $b$- and $c$-quark masses are, however, defined as the pole masses.

\begin{table}[t]
\caption{Relevant input parameters entering our numerical analysis. Details on the extraction of decay constants of vector mesons could be found in Refs.~\cite{Jung:2012vu,Ball:2006eu}.}
\label{tab:inputs}
\begin{tabular}{|l|l|}
\hline & \\[-0.6cm]
  $G_F = 1.1663787 \times 10^{-5}~\mathrm{GeV}^{-2}$   \hfill\cite{PDG2014}  &
  $f_{B_c} = 487 \pm 5~\mathrm{MeV}$                   \hfill\cite{Chiu:2007km} \\
  $m_b = 4.8 \pm 0.1~\mathrm{GeV}$              \hfill\cite{PDG2014} &
  $\tau_{B_c} = 0.452 \pm 0.033~\mathrm{ps}$           \hfill\cite{PDG2014}    \\
  $m_c = 1.5 \pm 0.1~\mathrm{GeV}$            \hfill\cite{PDG2014} &
  $f_{\pi} = 130.41 \pm 0.20~\mathrm{MeV}$             \hfill\cite{PDG2014}    \\
  $\overline{m}_s(2~\mathrm{GeV}) = 95 \pm 5~\mathrm{MeV}$        \hfill\cite{PDG2014} &
  $f_K = 156.1 \pm 0.8~\mathrm{MeV}$                   \hfill\cite{PDG2014}    \\
  $\overline{m}_s/\overline{m}_q = 27.5 \pm 1.0$                             \hfill\cite{PDG2014} &
  $f_{\rho} = 215\pm 6~\mathrm{MeV}$                   \hfill\cite{Jung:2012vu,Ball:2006eu}    \\
  $f_{\rho}^\perp(2~{\rm GeV})/f_{\rho} = 0.70\pm0.04$ \hfill\cite{Jung:2012vu,Ball:2006eu} &
  $f_{K^{*}} = 209\pm 7~\mathrm{MeV}$                   \hfill\cite{Jung:2012vu,Ball:2006eu}    \\
  $f_{K^{*}}^\perp(2~{\rm GeV})/f_{K^{*}} = 0.73\pm0.04$ \hfill\cite{Jung:2012vu,Ball:2006eu} &
  $f_{\omega} = 188\pm 10~\mathrm{MeV}$                   \hfill\cite{Jung:2012vu,Ball:2006eu}    \\
  $f_{\omega}^\perp(2~{\rm GeV})/f_{\omega} = 0.70\pm0.10$ \hfill\cite{Jung:2012vu,Ball:2006eu} &
  $f_{\phi} = 229\pm 3~\mathrm{MeV}$                   \hfill\cite{Jung:2012vu,Ball:2006eu}    \\
  $f_{\phi}^\perp(2~{\rm GeV})/f_{\phi} = 0.750\pm0.020$ \hfill\cite{Jung:2012vu,Ball:2006eu} &
  $m_g = 0.49 \pm 0.03{\rm GeV}$                         \hfill\cite{Chang:2008tf,Chang:2012xv}\\
[0.15cm]
\hline
\end{tabular}
\end{table}

\end{appendix}

\begin{table}[htbp]
\caption{The CP-averaged branching ratios~(in units of $10^{-8}$ for $|\Delta S|=0$ and $10^{-9}$ for $|\Delta S|=1$ transitions) of $B_{c}\to PP$~(upper) and $B_{c}\to PV$~(lower) decays based on the two schemes and three kinds of $B_{c}$ meson distribution function.}
\label{tab01}
\begin{tabular}{|c|c|c|c|c|c|c|c|}
  \hline\hline
  & &\multicolumn{2}{c|}{Wave-I}
  &\multicolumn{2}{c|}{Wave-II}
  &\multicolumn{2}{c|}{Wave-III}\\
  Decay modes &Cases &\quad S - I \quad
  &\quad S - II \quad
  &\quad S - I \quad
  &\quad S - II \quad
  &\quad S - I \quad
  &\quad S - II \quad\\
  \hline\hline
  $B_{c}^{-}\to\pi^{-}\pi^{0}$ & $|\Delta S|=0 $
  &0&0&0&0&0&0\\

  $B_{c}^{-}\to\pi^{-}\eta$ & $|\Delta S|=0 $
  &2.82&5.50&1.30&4.87&3.00&5.61\\

  $B_{c}^{-}\to\pi^{-}\eta^{\prime}$& $|\Delta S|=0$
  &1.86&3.63&0.86&3.21&1.98&3.70\\

  $B_{c}^{-}\to K^{-}{K}^{0}$& $|\Delta S|=0 $
  &4.69&9.15&2.18&9.11&5.01&9.25\\

  $B_{c}^{-}\to K^{-}\pi^{0}$& $|\Delta S|=1 $
  &0.92&1.79&0.43&1.86&0.98&1.81\\

  $B_{c}^{-}\to \bar{K}^{0}\pi^{-}$& $|\Delta S|=1 $
  &1.84&3.59&0.86&3.72&1.97&3.61\\

  $B_{c}^{-}\to K^{-}\eta$& $|\Delta S|=1 $
  &0.17&0.33&0.08&0.45&0.18&0.33\\

  $B_{c}^{-}\to K^{-}\eta^{\prime}$& $|\Delta S|=1 $
  &3.85&7.52&1.79&7.28&4.11&7.62\\ \hline\hline

  $B_{c}^{-}\to\pi^{-}\rho^{0}$ & $|\Delta S|=0$
  &$0$&$0$&$0$&$0.02$&0&$\sim0$\\

  $B_{c}^{-}\to \rho^{-}\pi^{0}$& $|\Delta S|=0$
  &$0$&$0$&$0$&$0.02$&0&$\sim0$\\

  $B_{c}^{-}\to\pi^{-}\omega$ & $|\Delta S|=0$
  &6.19&12.4&2.63&10.2&6.59&12.8\\

  $B_{c}^{-}\to \rho^{-}\eta$& $|\Delta S|=0 $
  &5.34&10.6&2.32&7.43&5.67&11.0\\

  $B_{c}^{-}\to \rho^{-}\eta^{\prime}$& $|\Delta S|=0 $
  &3.52&7.00&1.53&4.90&3.74&7.26\\

  $B_{c}^{-}\to K^{*-}{K}^{0}$& $|\Delta S|=0$
  &5.46&11.0&2.31&8.81&5.82&11.3\\

  $B_{c}^{-}\to K^{-}{K}^{*0}$& $|\Delta S|=0 $
  &5.46&11.0&2.31&9.94&5.82&11.3\\

  $B_{c}^{-}\to K^{-}\rho^{0}$& $|\Delta S|=1 $
  &1.53&3.07&0.65&2.30&1.63&3.17\\

  $B_{c}^{-}\to \bar{K}^{0}\rho^{-}$& $|\Delta S|=1 $
  &3.06&6.14&1.31&4.61&3.26&6.33\\

  $B_{c}^{-}\to K^{*-}\pi^{0}$& $|\Delta S|=1 $
  &1.04&2.09&0.44&1.98&1.11&2.14\\

  $B_{c}^{-}\to \bar{K}^{*0}\pi^{-}$& $|\Delta S|=1 $
  &2.07&4.19&0.87&3.95&2.21&4.28\\

  $B_{c}^{-}\to K^{-}\omega$& $|\Delta S|=1 $
  &1.17&2.35&0.50&1.77&1.25&2.42\\

  $B_{c}^{-}\to K^{*-}\eta$& $|\Delta S|=1 $
  &0.15&0.32&0.06&0.29&0.16&0.32\\

  $B_{c}^{-}\to K^{*-}\eta^{\prime}$& $|\Delta S|=1 $
  &4.58&9.22&1.95&7.41&4.88&9.47\\

  $B_{c}^{-}\to \phi K^{-}$& $|\Delta S|=1 $
  &3.55&7.18&1.49&7.02&3.78&7.33\\ \hline
\end{tabular}
\end{table}

\begin{table}[htbp]
 \caption{ The CP-averaged branching ratios and theoretical errors~(in units of $10^{-8}$) of $B_{c}\to PP(V)$ decays with $|\Delta S|=0$ based on W-I. The theoretical errors correspond to the uncertainties referred to as ``CKM", ``hadronic", ``scale", and ``$m_g$" defined in the text.}
 \label{tab02}
 \begin{tabular}{|l|c|c|c|}
 \hline\hline
 \multicolumn{1}{|c|}{Decay modes} & Cases & Scheme~I & Scheme~II  \\
 \hline\hline

 $B_{c}^{-}\to\pi^{-}\eta$ & $|\Delta S|=0 $
 & $2.82_{\,-0.20\,-1.94\,-2.03}^{\,+0.13\,+2.68\,+7.64}$
 & $5.50_{\,-0.39\,-3.77\,-3.16\,-0.12}^{\,+0.26\,+5.17\,+6.15\,+0.13}$
 \\

 $B_{c}^{-}\to\pi^{-}\eta^{\prime}$& $|\Delta S|=0$
 & $1.86_{\,-0.14\,-1.30\,-1.34}^{\,+0.08\,+1.81\,+5.04}$
 & $3.63_{\,-0.26\,-2.54\,-2.08\,-0.08}^{\,+0.16\,+3.48\,+4.05\,+0.08}$
 \\

 $B_{c}^{-}\to K^{-}{K}^{0}$& $|\Delta S|=0 $
 & $4.69_{\,-0.33\,-0.19\,-3.34}^{\,+0.22\,+0.19\,+12.24}$
 & $9.15_{\,-0.65\,-0.37\, -5.15\,-0.20}^{\,+0.43\,+0.38\,+9.70\,+0.21}$
\\
 $B_{c}^{-}\to\pi^{-}\omega$ & $|\Delta S|=0 $
 & $6.19_{\,-0.44\,-0.84\,-4.61}^{\,+0.29\,+0.92\,+19.23}$
 & $12.4_{\,-0.8\,-1.7\,-7.6\,-0.3}
 ^{\,+0.6\,+2.0\,+16.6\,+0.3}$
 \\

 $B_{c}^{-}\to \rho^{-}\eta$& $|\Delta S|=0 $
 & $5.34_{\,-0.38\,-0.74\,-3.98}^{\,+0.25\,+0.82\,+16.68}$
 & $10.6_{\,-0.7\,-1.8\,-6.5\,-0.2}
 ^{\,+0.5\,+2.1\,+14.3\,+0.3}$
 \\

 $B_{c}^{-}\to \rho^{-}\eta^{\prime}$& $|\Delta S|=0 $
 & $3.52_{\,-0.25\,-0.54\,-2.63}^{\,+0.16\,+0.60\,+10.99}$
 & $7.00_{\,-0.50\,-1.28\,-4.29\,-0.16}
 ^{\,+0.33\,+1.48\,+9.41\,+0.16}$
 \\

 $B_{c}^{-}\to K^{*-}{K}^{0}$& $|\Delta S|=0$
 & $5.46_{\,-0.39\,-0.54\,-4.07}^{\,+0.25\,+0.58\,+16.95}$
 & $11.0_{\,-0.8\,-1.1\,-6.7\,-0.3}
 ^{\,+0.5\,+1.2\,+14.6\,+0.3}$
 \\

 $B_{c}^{-}\to K^{-}{K}^{*0}$& $|\Delta S|=0 $
 & $5.46_{\,-0.39\,-0.54\,-4.07}^{\,+0.25\,+0.58\,+16.94}$
 & $11.0_{\,-0.8\,-1.1\,-6.7\,-0.3}
 ^{\,+0.5\,+1.2\,+14.6\,+0.3}$
 \\
 \hline\hline
\end{tabular}
\end{table}
\begin{table}[htbp]
 \caption{ Same as Table.\ref{tab02} but based on W-II.}
 \label{tab03}
 \begin{tabular}{|l|c|c|c|}
 \hline\hline
 \multicolumn{1}{|c|}{Decay modes} & Cases & Scheme~I & Scheme~II  \\
 \hline\hline

 $B_{c}^{-}\to\pi^{-}\eta$ & $|\Delta S|=0 $
 & $1.30_{\,-0.09\,-0.92\,-0.93}^{\,+0.06\,+1.41\,+3.48}$
 & $4.87_{\,-0.35\,-5.56\,-2.55\,-0.30}^{\,+0.22\,+11.31\,+4.16\,+0.39}$
 \\

 $B_{c}^{-}\to\pi^{-}\eta^{\prime}$& $|\Delta S|=0$
 & $0.86_{\,-0.06\,-0.62\,-0.62}^{\,+0.04\,+0.93\,+2.29}$
 & $3.21_{\,-0.23\,-3.71\,-1.68\,-0.20}^{\,+0.15\,+7.54\,+2.75\,-0.26}$
 \\

 $B_{c}^{-}\to K^{-}{K}^{0}$& $|\Delta S|=0 $
 & $2.18_{\,-0.15\,-0.08\,-1.55}^{\,+0.10\,+0.09\,+5.61}$
 & $9.11_{\,-0.64\,-0.36\, -4.61\,-0.63}^{\,+0.43\,+0.38\,+7.01\,+0.82}$
\\
 $B_{c}^{-}\to\pi^{-}\omega$ & $|\Delta S|=0 $
 & $2.63_{\,-0.18\,-0.34\,-1.96}^{\,+0.12\,+0.37\,+8.25}$
 & $10.2_{\,-0.7\,-1.9\,-6.1\,-1.1}
 ^{\,+0.5\,+2.3\,+12.5\,+0.3}$
 \\

 $B_{c}^{-}\to \rho^{-}\eta$& $|\Delta S|=0 $
 & $2.32_{\,-0.17\,-0.21\,-1.73}^{\,+0.10\,+0.21\,+7.27}$
 & $7.43_{\,-0.52\,-5.07\,-4.47\,-0.57}
 ^{\,+0.35\,+7.06\,+9.39\,+0.23}$
 \\

 $B_{c}^{-}\to \rho^{-}\eta^{\prime}$& $|\Delta S|=0 $
 & $1.53_{\,-0.11\,-0.16\,-1.14}^{\,+0.07\,+0.15\,+4.79}$
 & $4.90_{\,-0.35\,-3.42\,-1.94\,-0.38}
 ^{\,+0.23\,+4.75\,+6.18\,+0.15}$
 \\

 $B_{c}^{-}\to K^{*-}{K}^{0}$& $|\Delta S|=0$
 & $2.31_{\,-0.16\,-0.22\,-1.72}^{\,+0.11\,+0.24\,+7.24}$
 & $8.81_{\,-0.62\,-1.05\,-5.25\,-0.50}
 ^{\,+0.42\,+1.17\,+2.27\,+0.77}$
 \\

 $B_{c}^{-}\to K^{-}{K}^{*0}$& $|\Delta S|=0 $
 & $2.31_{\,-0.16\,-0.22\,-1.72}^{\,+0.11\,+0.24\,+7.24}$
 & $9.94_{\,-0.71\,-1.23\,-5.89\,-1.60}
 ^{\,+0.46\,+1.37\,+9.66\,+0.00}$
 \\
 \hline\hline
\end{tabular}
\end{table}

\begin{table}[htbp]
 \centering
 \tabcolsep 0.12in
 \caption{ \small Same as Table.\ref{tab02} but based on W-III.}
 \label{tab04}
 \begin{tabular}{|l|c|c|c|}
 \hline\hline
 \multicolumn{1}{|c|}{Decay modes} & Cases & Scheme~I & Scheme~II  \\
 \hline\hline

 $B_{c}^{-}\to\pi^{-}\eta$ & $|\Delta S|=0 $
 & $3.00_{\,-0.21\,-2.08\,-2.16}^{\,+0.14\,+2.91\,+8.13}$
 & $5.61_{\,-0.40\,-5.10\,-3.24\,-0.08}
 ^{\,+0.26\,+4.80\,+6.38\,+0.24}$
 \\

 $B_{c}^{-}\to\pi^{-}\eta^{\prime}$& $|\Delta S|=0$
 & $1.98_{\,-0.14\,-1.40\,-1.42}^{\,+0.09\,+1.96\,+5.36}$
 & $3.70_{\,-0.26\,-3.18\,-2.14\,-0.05}
 ^{\,+0.17\,+3.25\,+4.20\,+0.15}$
 \\

 $B_{c}^{-}\to K^{-}{K}^{0}$& $|\Delta S|=0 $
 & $5.01_{\,-0.36\,-0.20\,-3.57}^{\,+0.23\,+0.21\,+13.04}$
 & $9.25_{\,-0.66\,-0.37\, -5.25\,-0.13}
 ^{\,+0.43\,+0.39\,+10.01\,+0.37}$
\\
 $B_{c}^{-}\to\pi^{-}\omega$ & $|\Delta S|=0 $
 & $6.59_{\,-0.47\,-0.89\,-4.91}^{\,+0.31\,+0.99\,+20.46}$
 & $12.8_{\,-0.9\,-1.8\,-7.8\,-0.3}
 ^{\,+0.6\,+1.9\,+17.1\,+0.7}$
 \\

 $B_{c}^{-}\to \rho^{-}\eta$& $|\Delta S|=0 $
 & $5.67_{\,-0.40\,-0.82\,-4.23}^{\,+0.27\,+0.93\,+17.72}$
 & $11.0_{\,-0.8\,-1.4\,-6.7\,-0.1}
 ^{\,+0.5\,+1.8\,+14.9\,+0.6}$
 \\

 $B_{c}^{-}\to \rho^{-}\eta^{\prime}$& $|\Delta S|=0 $
 & $3.74_{\,-0.27\,-0.61\,-2.79}^{\,+0.17\,+0.67\,+11.67}$
 & $7.26_{\,-0.52\,-1.10\,-4.45\,-0.11}
 ^{\,+0.34\,+1.30\,+9.79\,+0.40}$
 \\

 $B_{c}^{-}\to K^{*-}{K}^{0}$& $|\Delta S|=0$
 & $5.82_{\,-0.41\,-0.58\,-4.34}^{\,+0.27\,+0.62\,+18.04}$
 & $11.3_{\,-0.8\,-1.1\,-6.9\,-0.2}
 ^{\,+0.5\,+1.2\,+15.1\,+0.6}$
 \\

 $B_{c}^{-}\to K^{-}{K}^{*0}$& $|\Delta S|=0 $
 & $5.82_{\,-0.41\,-0.58\,-4.34}^{\,+0.27\,+0.62\,+18.04}$
 & $11.3_{\,-0.8\,-1.1\,-6.9\,-0.2}
 ^{\,+0.5\,+1.2\,+15.1\,+0.6}$
 \\
 \hline\hline
\end{tabular}
\end{table}
\end{document}